\documentclass[letterpaper, 10 pt, conference]{ieeeconf}  

\IEEEoverridecommandlockouts                              
\usepackage[utf8]{inputenc}
\usepackage[T1]{fontenc}
\usepackage{graphicx}
\usepackage{amsmath}

\usepackage[ruled,vlined]{algorithm2e}

\usepackage{amsmath}
\usepackage{authblk}
\title{\LARGE \bf
Adaptive calibration of Heston Model using PCRLB based switching Filter
}
\author{Kumar Yashaswi$^{}$
}
\begin{document}
\maketitle
\thispagestyle{empty}
\pagestyle{empty}
\begin{abstract}

Stochastic volatility models have existed in Option pricing theory ever since the crash of 1987 which violated the Black-Scholes model assumption of constant volatility. Heston model is one such stochastic volatility model that is widely used for volatility estimation and option pricing. In this paper, we design a novel method to estimate parameters of Heston model under state-space representation using Bayesian filtering theory and Posterior Cramer-Rao Lower Bound (PCRLB), integrating it with Normal Maximum Likelihood Estimation (NMLE) proposed in [1]. Several Bayesian filters like Extended Kalman Filter (EKF), Unscented Kalman Filter (UKF), Particle Filter (PF) are used for latent state and parameter estimation. We employ a switching strategy proposed in [2] for adaptive state estimation of the non-linear, discrete-time state-space model (SSM) like Heston model. We use a particle filter approximated PCRLB [3] based performance measure to judge the best filter at each time step.
We test our proposed framework on pricing data from S\&P 500 and NSE Index, estimating the underlying volatility and parameters from the index. Our proposed method is compared with the VIX measure and historical volatility for both the indexes. The results indicate an effective framework for estimating volatility adaptively with changing market dynamics. 

\vspace{5mm} 

Keywords- Stochastic volatility; Heston Model; Normal MLE; Bayesian Filtering; Posterior Cramer-Rao Lower Bound (PCRLB)

\end{abstract}

Stochastic volatility models have been widely used for derivative pricing and risk management for various assets. These models are also used to judge the movement of asset prices and related factors. The market data is used widely to calibrate risk factors like volatility and drift. Ever since the violation of the constant volatility assumption assumed by Black-Scholes model, stochastic volatility models like Heston model have been widely used.
For equity modelling, Heston model is defined by constructing two related stochastic diffusion processes: the equity process and the volatility one. 
Heston model requires calibration of its parameters based on observed market data like asset prices, option prices and interest rates. A popular technique for calibration of parameters is Maximum Likelihood Estimation (MLE) along with non-linear Bayesian filtering for volatility estimation. 
 
[1] discussed the various limitation of using MLE to estimate parameters of Heston model. [4] gave a method to estimate the parameters of Heston model using the pseudo-MLE (PMLE) method, subject to known volatilities. This had the drawback of not being able to estimate correlation coefficient between the two diffusion processes and gave an approximated Gaussian distribution for the Heston model. To circumvent the problems arising in PMLE, [1] proposed normal Maximum Likelihood Estimation (NMLE) to estimate the parameters of Heston model. NMLE used transformation of Ito's lemma to get an exact Gaussian distribution of Heston model, as compared to approximate one. It also gave a method to estimate correlation coefficient using the moment estimation method. They tested to find NMLE a more efficient method than PMLE for estimating parameters. 
Both the maximum likelihood methods required known volatilities for estimation. They worked alternatively using maximum likelihood methods to estimate the parameters and filtering techniques to estimate volatility. For filtering technique Consistent Extended Kalman Filter (CEKF) was used, which is an extended version of EKF, bounding the estimation error in real-time. \par
The use of individual filters for estimation has the drawback that no individual filter will be optimal to use for the entire window of asset prices. The volatility may be more accurately modelled by different Bayesian filters in a given time period.

In this paper, we propose a framework for adaptive volatility and parameter estimation using switching Bayesian filter strategy based on Posterior Cramer-Rao Lower Bound (PCRLB) measure, which gives a theoretical lower bound for mean square error (MSE) of non-linear Bayesian filter state estimate.
[2] proposed an efficient technique to select the most optimal filter at each time step from a set of Bayesian filters for a general non-linear, stochastic, discrete-time state-space model (SSMs). 
The switching strategy switches between a set of Bayesian filters at each time step based on their error covariance and PCRLB measure. Similar technique was used in our work in [5], applied to an alternative problem of option price forecasting over a Black-Scholes model framework. \par
We estimate the volatility using PCRLB based switching filter and correspondingly calibrating Heston model parameters using NMLE method. Computing PCRLB is done using a particle filter (PF) based approximation based on works of [3]. 
The use of adaptive state estimation strategy and NMLE gives a stable framework for estimating the underlying volatility and parameters. We analyse the performance of volatility estimation on $S\&P$ 500 and National Stock Exchange (NSE) prices under Heston model formulation.

The main aim of our work is to investigate the efficiency of using the proposed filter switching strategy for adaptive volatility and parameter estimation of Heston model. Such a framework can be used to estimate volatility movement of any asset and further used in option pricing under an appropriate modelling framework.

\subsection{Related Work}
Having used similar framework as [5], we wanted to explore the application and performance on calibration of stochastic volatility models. Additionally, [1,4,6] provided a framework for parameter estimation and volatility estimation using Bayesian filtering. The state-space representation and usage of filtering theory was aided by their proposed method.
[7,8,9,10] gives a basic overview of filtering theory in finance including in Stochastic Volatility model. [9] further extended the works to delta hedging after estimating parameters of Stochastic volatility models. \par

\subsection{Problem Formulation}
We study equity price modelling by Heston model under risk-neutral measure $Q$, and formulate it in the dynamics of state-space model. The model assumes the underlying stock price $S_t$ has a stochastic variance $V_t$ following a CIR process. The dynamics are represented as:
\begin{equation}
dS_t=rS(t)dt + \sqrt{V_t} S(t)dW_1(t)
\end{equation}
\begin{equation}
dV_t=\kappa(\theta - V_t)dt + \sigma\sqrt{(V(t))} dW_2(t)
\end{equation}
$W_1(t)$ and $W_2(t)$ are two Brownian motions correlated with a constant correlation $\rho$ such that:
\begin{equation}
(dW_1(t),dW_2(t))= \rho dt, t \in [0,T]
\end{equation}
$r$ is a constant risk-free interest rate.
The mean-reverting process is the term $\kappa(\theta - V_t)$, with $\theta$ is the long-run mean reversion level of the volatility, $\kappa$ is the mean
reversion rate towards $\theta$ and $\sigma$ is a volatility of the volatility. All the parameters $\kappa$, $\theta$, $\sigma$ are positive constant. Positiveness of the volatility is ensured when $2\kappa\theta > \sigma^2$. \par 
Heston model can be represented in the form of state-space filtering problem formulation. Since volatility is an unknown value, Eq (1-2) can be regarded as measurement and state equation. Continuous form of the model can be written as- 

\begin{equation}
d V(t)= \kappa(\theta - V(t))dt + \sigma \sqrt{V(t)}dW_v
\end{equation}
\begin{equation}
\begin{split}
d \ln(S(t))= (r - \frac{1}{2}V(t))dt + \sqrt{(1 - \rho^2)V(t)}dW_s(t) \\
+ \rho\sqrt{V(t)}dW_v(t), t\in [0,T]
\end{split}
\end{equation}
where $dW_s(t)dW_v(t)=0$, and errors are transformed as: 

\begin{equation}
W_1(t) = \sqrt{(1 - \rho^2)}dW_s(t) + \rho W_v(t)
\end{equation}
\begin{equation}
W_2(t) = W_v(t)
\end{equation}
The parameters of Heston model $(\kappa,\theta,\sigma,\rho)$ are unknown. The volatilities are estimated using filtering techniques that require known parameters. As discussed in [1,4] parameter estimation and volatility tracking is a coupled issue that is done synchronously. One major addition to their algorithm is adding PCRLB based switching filter after each step of parameter estimation.
[6] proposed an alternate representation of Heston model with some rearranging. 
From the observed stock price $S(t)$, we form the transformed observation process $y_t = log S(t)/S(0)$:
\begin{equation}
d y_t= (r - \frac{1}{2}V_t)dt + \sqrt{V_t}dW_1(t)
\end{equation}
Taking the observations ${\{y_s\}}_{0\leq s \leq t}$, we set:
\begin{equation}
\Tilde{W(t)}= \frac{1}{\sqrt{1 - \rho^2}}(W_2(t) - \rho W_1(t))
\end{equation}
where $\Tilde{W(t)}$ is independent of $W_1(t)$. Noting that
\begin{equation}
\begin{split}
d W_2(t)= \sqrt{1 - \rho^2}d\Tilde{W(t)} + \rho d W_1(t) \\
= \sqrt{1 - \rho^2}d\Tilde{W(t)} + \frac{\rho}{\sqrt{V_t}}(dy_t - (r - \frac{1}{2}V_t)dt)
\end{split}
\end{equation}
we have
\begin{equation}
\begin{split}
d V_t= \kappa(\theta - V_t)dt + \sigma \sqrt{V_t} \sqrt{1 - \rho^2}d\Tilde{W(t)} \\ 
+ \sigma \rho (dy_t - (r - \frac{1}{2}V_t)dt)
\end{split}
\end{equation}
Discretizing the equations we get:
\begin{equation}
\begin{split}
V_k= V_{k-1} + \kappa(\theta - V_{k-1})\triangle t - \sigma \rho(r - \frac{1}{2}V_{k-1})\triangle t \\ 
+ \sigma \sqrt{V_{k-1}} \sqrt{1 - \rho^2} \triangle\Tilde{W(t)} + \sigma \rho(y_k - y_{k-1})
\end{split}
\end{equation}

\begin{equation}
y_k= y_{k-1} + (r - \frac{1}{2}V_{k})\triangle t + \sqrt{V_{k-1}}\triangle W_1(t)
\end{equation}

\section{Background}

\subsection{Normal Maximum Likelihood Estimation} 
We discuss the likelihood estimation method as used originally in [1], which was an alternative to Pseudo-Maximum likelihood estimation used in [4]. They assumed the parameters of the diffusion process as fixed coefficients, hence a recursive form of volatility was approximated with Gaussian process. However, [1] argued that conditioned on previous value V(t - $\delta$) ($\delta$ is an arbitrary small positive number), the distribution of V(t) is noncentral chi-square with non centrality paramerter $nc = \frac{4 \kappa e^{-\kappa \delta}}{\sigma^2(1 - e^{-\kappa\delta})}V(t - \delta)$ and degrees of freedom $df = \frac{4\kappa\theta}{\sigma^2}$. They more appropriately modelled volatility as exact Gaussian distribution. Since volatility follows chi-square distribution, it's intuitive that the square root of volatility follows Gaussian distribution. \par
Without jumping into the derivation, the parameters are estimated as:
\begin{equation} 
\widehat{\kappa} = \frac{2}{\delta}(1+ \frac{\widehat{P}\delta}{2}\frac{1}{n}\sum_{k=1}^{n}\frac{1}{V_{k-1}} - \frac{1}{n}\sum_{k=1}^{n}\sqrt{\frac{V_{k}}{V_{k-1}}})
\end{equation}

\begin{equation} 
\widehat{\sigma} = \sqrt{\frac{4}{\delta}\frac{1}{n}\sum_{k=1}^{n}[\sqrt{V_{k}} - \sqrt{V_{k-1}} - \frac{\delta}{2\sqrt{V_{k-1}}} (\widehat{P} - \widehat{\kappa}V_{k-1})]^2}
\end{equation}

\begin{equation} 
\widehat{\theta} = \frac{\widehat{P} + \frac{1}{4}\widehat{\sigma}^{2}}{\widehat{\kappa}}
\end{equation}
where:
\begin{equation} 
\widehat{P} = \frac{\frac{1}{n} \sum_{k=1}^{n} \sqrt{V_{k-1} V_{k}} - \frac{1}{n^2}\sum^{n}_{k=1} \sqrt{\frac{V_k}{V_{k-1}}}\sum_{k=1}^{n} V_{k-1}}{\frac{\delta}{2} - \frac{\delta}{2} \frac{1}{n^2} \sum_{k=1}^{n}\frac{1}{V_{k-1}} \sum_{k=1}^{n} V_{k-1}} 
\end{equation}
NMLE gives another advantage over PMLE that helps in direct estimation of correlation coefficient $\rho$ as:
\begin{equation} 
\widehat{\rho} = \frac{1}{n\delta} \sum_{k=1}^n \triangle W_{1_k} \triangle W_{2_k}
\end{equation}

for the discrete form as:
\begin{equation} 
\triangle W_{1_k} = \frac{\ln S_k - \ln S_{k-1} - (r - \frac{1}{2}V_{k-1})\delta}{\sqrt{V_{k-1}}}
\end{equation}

\begin{equation} 
\triangle W_{2_k} = \frac{V_k - V_{k-1} - \kappa(\theta - V_{k_1})\delta}{\sigma\sqrt{V_{k-1}}}
\end{equation}

where $\triangle W_{i_k} = W_{i_k} - W_{i_{k-1}}$ for $i=1,2,..$.

\subsection{Posterior Cramér-Rao lower bound}
Cramer–Rao bound (CRB) is a commonly used statistical performance bound to indicate the lower theoretical bound of the MSE of a state-estimator. The particle filter approximation algorithm is explained in detail in [3] and [5]. We provide an overview in this paper describing the result of the algorithm.
In time-varying systems such as the asset-price movement, the estimated price movement is considered random since it corresponds to an underlying nonlinear, randomly driven state process. For such systems a Posterior Cramer–Rao bound (PCRLB) is computed. 
For a system of non-linear model:
\begin{equation}
X_{t+1} = f_{t}(X_t)+V_t,\\
\hspace{0.1cm} V_t \sim N(0,Q_t) 
\end{equation}
\begin{equation}
Y_{t} = g_{t}(X_t)+W_t,\\
\hspace{0.1cm} W_t \sim N(0,R_t)
\end{equation}
the PCRLB inequality is described as:
\begin{equation}
P_{t|t} \equiv E_{p(X_{0:t},Y_{1:t})}[(X_t - {\widehat{X}}_{t|t})(X_t - {\widehat{X}}_{t|t})^T] \geq J_t^{-1}
\end{equation}
which can also be written in the form 
\begin{equation}
P_{t|t}^S \equiv E_{p(X_{0:t},Y_{1:t})}[\|X_t - {\widehat{X}}_{t|t}\|^2] \geq Tr[J_t^{-1}]
\end{equation}
Eq (24) can be attributed to the fact that $P_{t|t}-J_t^{-1}\geq0$ is a positive definite matrix for all state estimates at time $t\in N$.
The model in Eq (21-23), following set of assumptions (Assumption (3.2-3.5) as described in [2]), gives way to the recursive computation of the PFIM as: 
\begin{equation}
J_{t+1} = D_t^{22} - [D_t^{12}]^T (J_{t}+D_t^{11})^{-1}D_t^{12}
\end{equation}
The individual components are defined as in Eq (26-28) below:
\begin{equation}
D_t^{11} = E_{p(X_{0:t+1},Y_{1:t+1})}[-\triangle_{X_{t}}^{X_{t}} log(p(X_{t+1}|X_t))]
\end{equation}
\begin{equation}
D_t^{12} = E_{p(X_{0:t+1},Y_{1:t+1})}[-\triangle_{X_{t}}^{X_{t+1}} log(p(X_{t+1}|X_t))]
\end{equation}
\begin{equation}
\begin{split} 
D_t^{22} = E_{p(X_{0:t+1},Y_{1:t+1})}[-\triangle_{X_{t+1}}^{X_{t+1}} log(p(X_{t+1}|X_t)) \\- \triangle_{X_{t+1}}^{X_{t+1}} log(p(Y_{t+1}|X_{t+1}))]
\end{split}
\end{equation}

The gradients using the Laplacian operator $\triangle$ is calculated at the true
states. The PFIM matrix is defined for initial states as:
\begin{equation}
J_{0} = E_{p(X_{0})}[-\triangle_{X_{0}}^{X_{0}} log(p(X_{0})]
\end{equation}

\subsection{Particle Filter Approximation of PCRLB}
Since Heston is a  non-linear SSM with additive Gaussian noise, we work with the model represented by Model 4.15 in [2]. Though the errors in Heston model are Gaussian but not directly additive, we still work with the particle filter based approximation model. 
Based on the approach mentioned in [2], approximation of Eq (26-28) was derived as:
\begin{equation}
D_t^{11} = \frac{1}{MN}\sum_{j=1}^{M}\sum_{i=1}^{N}[\nabla_{X_{t}} f_{t}^T(X_{t|t+1}^{i,j})]Q_{t}^{-1}[\nabla_{X_{t}} f_{t}(X_{t|t+1}^{i,j})]
\end{equation}
\begin{equation}
D_t^{12} = \frac{1}{MN}\sum_{j=1}^{M}\sum_{i=1}^{N}-[\nabla_{X_{t}} f_{t}^T(X_{t|t+1}^{i,j})]Q_{t}^{-1}
\end{equation}
\begin{equation}
\begin{split} 
D_t^{22} = Q_{t}^{-1} + \frac{1}{MN}\sum_{j=1}^{M}\sum_{i=1}^{N}([\nabla_{X_{t+1}} g_{t+1}^T(X_{t+1|t}^{i,j})]\\ R_{t+1}^{-1}[\nabla_{X_{t+1}} g_{t+1}(X_{t+1|t}^{i,j})]) 
\end{split}
\end{equation}
The particles are N-samples following $\{X_{t+1|t}^{i,j}\}_{i=1}^{N} \sim p(x_{t+1}|y_{1:t}^j)$ and $\{X_{t|t+1}^{i,j}\}_{i=1}^{N} \sim p(x_{t}|y_{1:t+1}^j)$, with M measurement sequence being $\{Y_{1:t+1}=y_{1:t+1}\}_{i=1}^{M}$ obtained from historical data. For the problem of volatility estimation there is only 1 measurement sequence. \par

The particle filter approximation of both PFIM and PCRLB using Eq (30-32) and matrix inversion lemma are given by:
\begin{equation}
J_{t} = D_t^{22} - [D_t^{12}]^T (J_{t}+D_t^{11})^{-1}D_t^{12}
\end{equation}
\begin{equation}
\begin{split}
J_{t}^{-1} = [D_t^{22}]^{-1} - [D_t^{22}]^{-1}[D_t^{12}]^T \times [D_t^{12}[D_t^{22}]^{-1} [D_t^{12}]^{T} \\ - (J_{t}+D_t^{11})]^{-1} D_t^{12}[D_t^{22}]^{-1}
\end{split}
\end{equation}
where $J_{t}$ and $J_{t}^{-1}$ are particle filter approximated PFIM and PCRLB respectively.

\subsection{Adaptive State-Estimation Switching Strategy}
PCRLB judges the MSE performance of a Bayesian estimator, with the estimator with the closest MSE to PCRLB being the best performing. MSE matrix is computed for each filter and PCRLB through particle filter approximation explained in the last subsection. Based on this, the strategy is initiated with a set of Bayesian estimators, and switch between them as and when
required based on their performance. To get a scalable metric we define performance metric as:
\begin{equation}
\Phi_t = J_t^{-1}oP_{t|t}^{-1}
\end{equation}
$\Phi_t \in R^{s\times s}$ is the new performance matrix to assess the performance of the Bayesian state estimator used, where $s$ is the number of states to be estimated.  An alternative definition of equation (35) is:
\[
\Phi_t (i,j) =
\begin{cases}
J_t(j,j)^{-1}[P_{t|t}(j,j)]^{-1}, &\text{if i=j;}\\
0, &\text{if i $\neq$ j.}\\
\end{cases}
\]
where $i,j \in 1,2,..s$. We state some results from [1] with given proof in the same-
\begin{equation}
0< \Phi_t(j,j) \leq 1
\end{equation}

This lead's to the condition $Trace[\Phi_t] \in (0,s]$ for all Bayesian estimators at any time. This puts a bound on the performance measure as compared to PCRLB which may be unbounded in many applications.
For the strategy, the best point estimate at each time step is decided based on the maximum value of performance metric for each state estimator $f\in B$. The process is initiated by computing the performance metrics using MMSE and PCRLB matrix for all Bayesian estimators from $B$. We then switch between them for each time step based on which gave the best performance. We employ the Average case switching strategy from [1] for our state estimation. The average-case strategy can be described as:
\begin{itemize}
\item Input: Set of measurements up to time period t, set of Bayesian filters $B$
\item Step 1: Compute MMSE matrix, PCRLB and Performance metric for all Bayesian estimator $f\in B$
\item Step 2: Compute $Trace[\Phi_t^f]$ using Performance metric for all $f\in B$
\item Step 3: Solve for f such that $\widehat{f} = arg max_{f}(Trace[\Phi_t^f])$ 
\item Output: $\widehat{X}_{t|t}^{(\widehat{f})}$ is the point estimate to be selected at time $t$ 
\end{itemize}
  
This technique also has the advantage of parallel computation for increasing computation of the sub-optimal state. All the estimators in $B$ are used at each sampling time parallelly, but only the one suggested by the proposed strategy is selected for the final estimate. Boundedness of stochastic processes is used to show the numerical stability of the algorithm [2].

\begin{figure*}[h]
\includegraphics[width=\textwidth, height=3 cm]{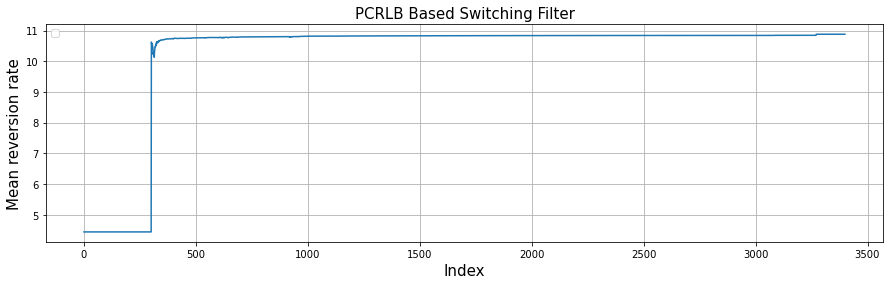}
\vspace*{\floatsep}
\includegraphics[width=\textwidth, height=3 cm]{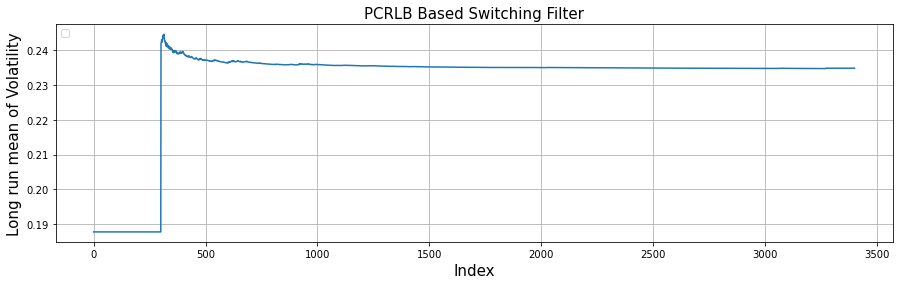}
\vspace*{\floatsep}
\includegraphics[width=\textwidth, height=3 cm]{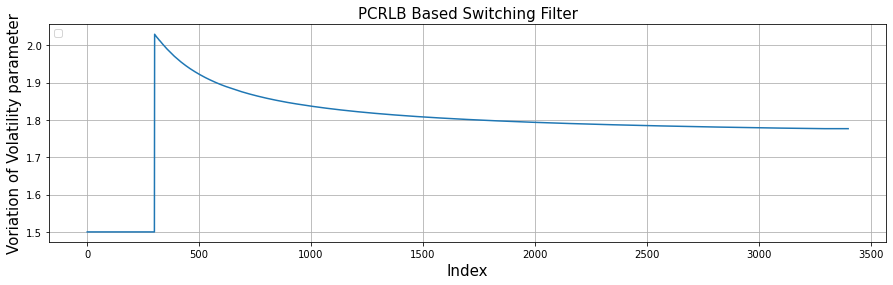}
\caption{Heston model Parameters estimated through the proposed methods}
\centering
\end{figure*}

\begin{figure*}[h]
\includegraphics[width=\textwidth, height=4 cm]{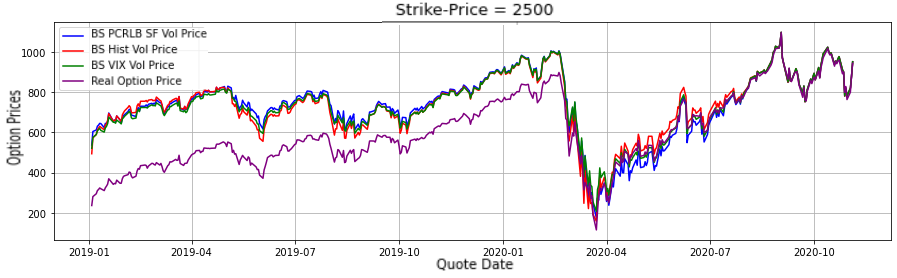}
\caption{Performance on pricing options}
\centering
\end{figure*} 

\begin{figure*}[h]
\includegraphics[width=\textwidth, height=4 cm]{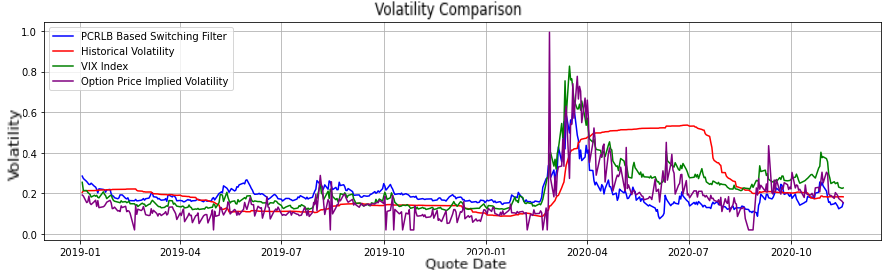}
\caption{Volatility Comparison with Implied Volatility}
\centering
\end{figure*} 

\begin{figure*}[h]
\includegraphics[width=\textwidth, height=4 cm]{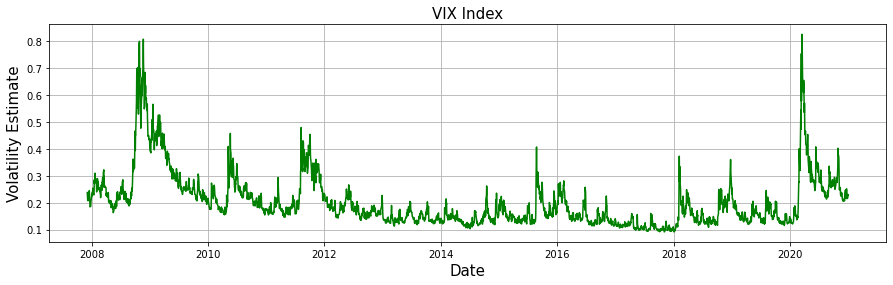}
\vspace*{\floatsep}
\includegraphics[width=\textwidth, height=4 cm]{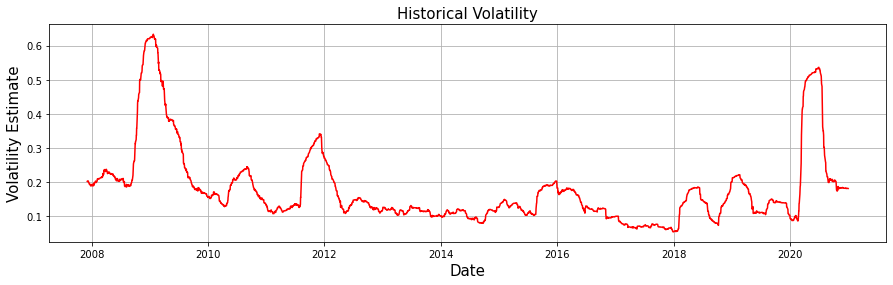}
\vspace*{\floatsep}
\includegraphics[width=\textwidth, height=4 cm]{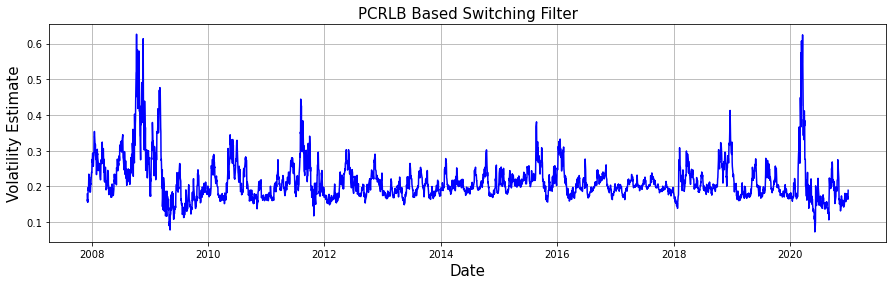}
\caption{Volatility Comparison of the switching filter with VIX and Historical Volatility- S\&P 500 Index}
\centering
\end{figure*} 

\section{Data Used}

In the previous sections, we have described our proposed strategy based on PCRLB measure and combining it with NMLE to estimate volatility under Heston model framework. We use 3 different filters (EKF, UKF, PF) in the Bayesian filters set and particle filter for approximate computation of PCRLB. 

The data used for testing for the proposed method are the S$\&$P 500 price data from the year 2007 to 2020 which gives a good indication of the U.S stock market. For the risk-free rate we use yield on long term US-Treasury. Similarly, the method is tested on NSE index data from 2012 to 2021, which gives an indication of the Indian stock market. This was done to test the results on those markets where quantitative tools have not been used widely.  \par
We compare the estimated volatilities tracked by PCRLB based method with VIX index and historical volatilities during the same period. While VIX reflects the future expectation of the volatilities, the historical volatilities reflect the past. They are used as reference measures. We additionally consider S$\&$P 500 options data from 2019-2020, and estimate the performance of volatility on pricing options under a Black-Scholes pricing framework. A performance comparison is done with respect to close-of-day implied volatility measure obtained from most liquid options contracts traded each day. The options data was taken from CBOE [11].

\begin{figure*}[h]
\includegraphics[width=\textwidth, height=4 cm]{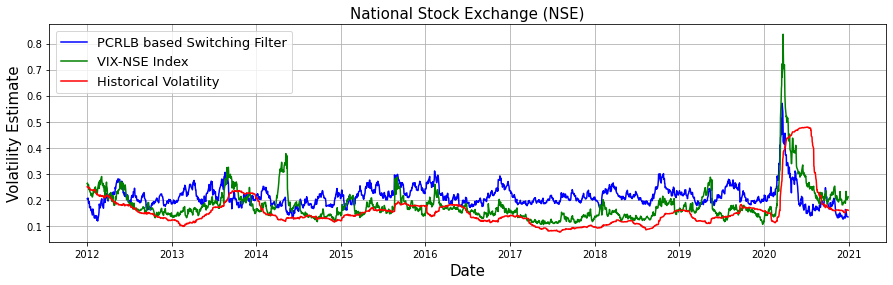}
\caption{Volatility Comparison of the switching filter with VIX and Historical Volatility- NSE Index}
\centering
\end{figure*}
\section{Experimental Results and Analysis}
Fig (1) shows the parameters of Heston model determined by the proposed method. Beginning with fixed arbitrary values in a given range for initial few steps, the algorithm converges to a stable value of the parameters soon as the method is added. The final parameter value was as: $[\kappa,\theta,\sigma]=[10.8, 0.23, 1.78]$, keeping correlation coefficient fixed at 0.8. Fig (4) shows the volatility structure obtained by these parameters and compared to different methods. The proposed method replicates the VIX Index and Historical volatility well, using only the data from index prices. The values are roughly between VIX index and historical volatility indicating the technique captures both historical and forward looking components of the market. Though for most of the time period particle filter was used, wherever needed the strategy switched to UKF and EKF for better capturing the market.\par

The performance on an option contract of strike price 2500, is not good initially, but after a fall in option price during March 2020, it manages to replicate the option prices.

Table (1) displays the distance between the estimated prices and real option prices, with all 3 estimated volatility performing similarly with Black-Scholes pricing framework. 
We compare the volatilities with the implied volatility for most traded options per day during the two year period. The PCRLB based switching filter estimated volatility represents VIX index better than implied volatility measure.  \par
The results on NSE index are similar, where it successfully replicates the VIX NSE index, though quantitative tools are not so widely used for NSE index yet. The volatility estimated is relatively stable in values as compared to VIX-NSE index.
The final parameters estimated for Heston model for NSE index data is as: $[\kappa,\theta,\sigma]=[7.5, 0.24, 1.13]$.  
\begin{table}
\begin{center}
\begin{tabular}[t]{ |p{1.5cm}||p{1.5cm}|p{1.5cm}|p{1.5cm}| }
 \hline
 \multicolumn{4}{|c|}{RMSE}\\
 \hline
 & VIX Index & Historical Vol & Proposed Method\\
 \hline
Real Option Prices   & 0.065    & 0.069 &  0.067\\
 \hline
\end{tabular}
\end{center}
\caption{RMSE for Price Forecasting on Option with k=2500}
\label{table:ta}
\end{table} 
\section{CONCLUSIONS and FUTURE WORK}

In this paper, we propose using PCRLB based switching filter for calibrating volatility and parameter of Heston model. We test the method on S\&P 500 and National Stock Exchange (NSE) index data, estimating a volatility structure similar to VIX data and estimating the parameter, which converges to a suitable, stable value.  
The method has the benefits of using a combination of filters, adapting to changing market dynamics. The technique has application to many other sub-fields of finance like term structure of commodity futures and interest rates, price prediction of stock's, exotic option's or other complex derivatives, subject to good modelling framework.



\end{document}